%% file: main.tex
\documentclass[
  aps,
  prx,          
  reprint,
  amsmath,amssymb,
  floatfix,
]{revtex4-2}

\usepackage{graphicx}
\usepackage{hyperref}
\usepackage{siunitx}
\usepackage{xcolor}

\DeclareSIUnit{\angstrom}{\text{\AA}}

\graphicspath{{figures/}}

\begin{document}

\title{Conformational landscape of a macrocycle using enhanced sampling}

\author{Valentin Kasper}
\email{valentin@pexmachina.com}
\author{Nicole Holzmann}
\author{Sanjoy Ray}
\author{Matthias Kaiser}
\affiliation{PexMachina Inc., 8 The Green, STE R, Dover, DE 19901, USA}
\author{Alons Lends}
\affiliation{National Institute of Research and Innovation, Aizkraukles 21, Riga, LV-1006, Latvia}

\date{\today}

\begin{abstract}
\input{sections/abstract}
\end{abstract}

\maketitle

\input{sections/introduction}

\input{sections/methods}

\input{sections/results}
\input{sections/discussion}

\input{sections/conclusion}
\input{sections/acknowledgments}

\bibliography{refs}

\end{document}

%% file: sections/abstract.tex
We use replica-exchange molecular dynamics (REMD) to map the conformational free-energy landscape of the macrocyclic drug lorlatinib in explicit water and chloroform.
In water, a single dominant basin is recovered; in chloroform, two conformational states are resolved.
These conformers serve as starting points to calculate proton chemical shifts from first principles within a continuum solvent model.
The resulting population-weighted spectrum is compared directly with the experimental CDCl$_3$ spectrum, which benchmarks the computational approach against measured NMR data.

%% file: sections/introduction.tex
\section{Introduction}

\begin{figure}[htp]
  \centering
  \includegraphics[width=0.55\columnwidth]{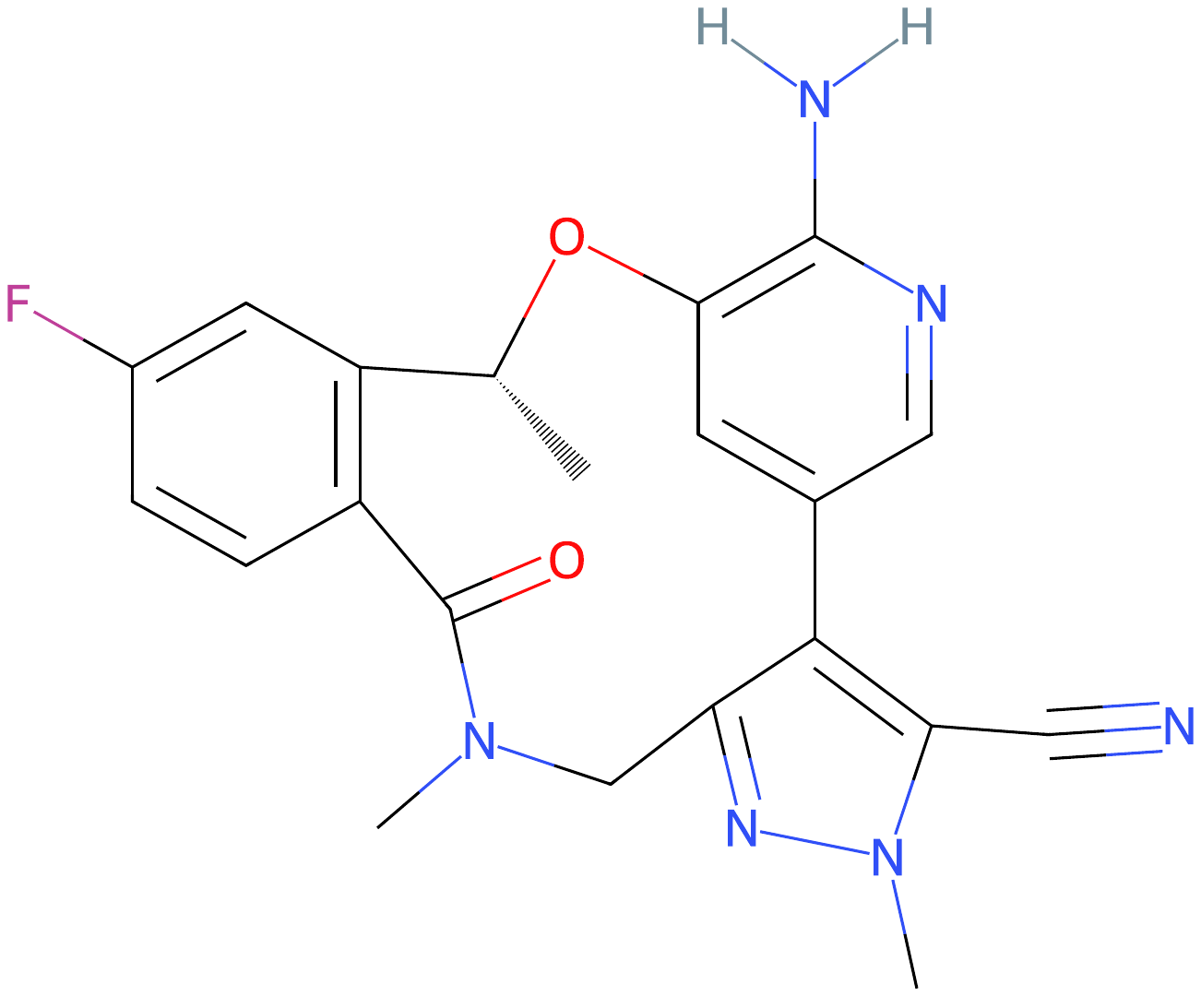}
  \caption{%
    \textbf{Chemical structure of lorlatinib} (C$_{21}$H$_{19}$FN$_6$O$_2$, MW 406.4~Da).
    The 17-membered macrocyclic ring is closed by a benzylic ether and an $N$-methyl amide bond, bridging the fluorinated benzene ring (left) to the 1-methylpyrazole-pyridine biaryl (right) and constraining the relative orientation of the two heterocycles.
    A nitrile on the pyrazole C3 position and an aminopyridine group project away from the ring and form the kinase hinge-binding pharmacophore; the hashed wedge bond at the benzylic carbon indicates the single fixed stereocenter.
    \label{fig:lorlatinib}}
\end{figure}

Macrocycles are an unusual class of drugs: large enough to engage flat or otherwise intractable protein surfaces, yet often orally bioavailable~\cite{driggers2008macrocycles,giordanetto2014macrocyclic}.
This combination is largely attributed to their conformational flexibility.
Many macrocyclic drugs can fold to bury their polar groups in nonpolar environments and expose them again for target binding, a chameleonic property that arises from the coupling of torsional degrees of freedom around the ring~\cite{whitty2016chameleonic}.
Lorlatinib (Fig.~\ref{fig:lorlatinib}), the approved ALK/ROS1 inhibitor, exemplifies this class: it was designed as a macrocyclic analogue of crizotinib, with the ring closure improving metabolic stability, brain penetration, and activity against resistance mutations~\cite{johnson2014discovery}.

Peng et al.\ showed by NMR and MD that lorlatinib populates two distinct conformers in chloroform (P1, major, $\approx$80\%; P2, minor, $\approx$20\%) separated by a barrier of approximately 22~kJ/mol, while only P1 is detectably populated in water~\cite{peng2019lorlatinib}.
This solvent-dependent equilibrium is directly relevant to lorlatinib's pharmacokinetics: the minor conformer is thought to present a less polar face in membrane-like environments, while P1 is the bioactive conformation at the kinase binding site.
Characterizing this balance computationally, and validating the result against experiment, is the goal of this work.

Solution NMR provides the primary experimental handle on the conformational equilibrium.
The observed $^1$H chemical shift is a population-weighted average over all rapidly interconverting states, so the one-dimensional spectrum encodes the conformer balance directly.
Two-dimensional NOESY spectra complement it through cross-peaks between protons that are spatially close regardless of their connectivity, with intensities that fall off as $r^{-6}$~\cite{solomon1955relaxation}.
Together the $^1$H spectrum and NOESY cross-peak pattern form a conformational fingerprint against which any simulation can be tested.

Molecular dynamics (MD) simulation offers a computational route to the same conformational ensembles that NMR probes experimentally.
A single unbiased trajectory is rarely long enough to cross the torsional barriers separating macrocycle conformers, so enhanced-sampling methods are required.
Temperature replica-exchange MD (T-REMD)~\cite{swendsen1986,hukushima1996,marinari1992,sugita1999replica} runs many copies of the system at a ladder of temperatures; high-temperature replicas cross barriers freely, and periodic coordinate swaps between neighboring replicas preserve the correct canonical distribution at the reference temperature while accelerating barrier crossing~\cite{earl2005}.

The accuracy of any such simulation depends on the underlying force field.
Classical, additive force fields such as GAFF~\cite{wang2004gaff} and OpenFF Sage~\cite{openff2022sage} assign fixed atomic charges and independently fitted torsional parameters, which gives them no mechanism to represent how ring closure couples torsions or introduces unusual transannular contacts.
A recent benchmark against experimental NMR restraints for a panel of macrocycles found substantial force-field-dependent variation, with even the best classical descriptions showing systematic deviations for some compounds~\cite{waibl2024validating}.

We report T-REMD simulations of lorlatinib in explicit water and chloroform with the classical OpenFF Sage force field and validate the resulting free-energy landscapes against experimental $^1$H NMR spectra and NOESY cross-peak patterns recorded in CDCl$_3$.

%% file: sections/methods.tex
\section{Methods}

This section describes the system setup, the molecular dynamics protocol, the enhanced-sampling strategy, and the NMR observables computed from the resulting conformational ensemble.

\subsection{System preparation}
\label{sec:system}

Lorlatinib was constructed from its SMILES string and a single conformer was generated with the OpenFF toolkit.
The ligand was parameterized with the OpenFF 2.0.0 force field (\texttt{openff-2.0.0.offxml})~\cite{openff2022sage}.
Partial charges for the chloroform simulations were assigned using the NAGL graph neural network charge model (\texttt{openff-gnn-am1bcc-0.1.0-rc.3.pt}); for the water simulations charges were assigned via the default AM1-BCC procedure of the OpenFF \texttt{SystemGenerator}.

\subsubsection{Chloroform system}
Explicit chloroform was represented as individual CHCl$_3$ molecules parametrized with the OpenFF 2.0.0 force field using the same NAGL charge model as the solute.
The simulation box was packed with Packmol using 150 CHCl$_3$ molecules at the experimental density of 1.48~g\,mL$^{-1}$ at 25\textdegree C, producing a box of $3.07 \times 3.07 \times 2.17$~nm containing 799 particles in total.

\subsubsection{Aqueous system}
The aqueous system was built with the \texttt{SystemGenerator} from \texttt{openmmforcefields}, combining the AMBER TIP3P water model (\texttt{amber/tip3p\_standard.xml}) with the OpenFF Sage 2.0.0 small-molecule parameters (\texttt{openff-2.0.0.offxml}).
The solvation box was constructed with OpenMM's \texttt{Modeller.addSolvent} using 1.0~nm of padding on all sides and an ionic strength of 0.15~M NaCl.

\subsubsection{Energy minimization and equilibration}
Both systems were energy-minimized with the OpenMM L-BFGS minimizer, then equilibrated in two stages using a Langevin middle integrator with a 2~fs time step and a friction coefficient of 1~ps$^{-1}$.
First, 100~ps of NVT dynamics at 300~K were run to relax the solvent shell around the solute.
Second, 1~ns of NPT dynamics at 300~K and 1~bar (Monte Carlo barostat) was run to equilibrate the box volume.
All simulations were run on a single NVIDIA GPU.
\subsection{Molecular dynamics}
\label{sec:md}

All simulation systems (water, chloroform) were propagated under periodic boundary conditions with particle mesh Ewald (PME) summation for long-range electrostatics and a direct-space cutoff.
Production sampling was conducted in the NVT ensemble at fixed box volume, with periodic box vectors inherited from the final frame of the NPT equilibration.

\subsection{T-REMD sampling}
\label{sec:rest}

Macrocycles populate multiple conformational states separated by barriers of several $k_BT$, making standard MD inefficient: trajectories become trapped and underestimate the weight of minority conformers on accessible timescales.
Temperature replica-exchange MD (T-REMD)~\cite{swendsen1986,hukushima1996,marinari1992,sugita1999replica} addresses this by running $M$ copies of the system in parallel at a ladder of temperatures $T_1 < T_2 < \cdots < T_M$.
High-temperature replicas cross barriers freely; periodically, adjacent replicas attempt a coordinate swap accepted with probability
\begin{equation}
  \label{eq:remd}
  P_{m \leftrightarrow m+1}
  = \min\!\left(1,\, e^{(\beta_m - \beta_{m+1})(U_m - U_{m+1})}\right),
\end{equation}
where $\beta_m = (k_B T_m)^{-1}$ and $U_m$ is the potential energy of replica $m$.
Swapping injects high-energy configurations into the cold ensemble, accelerating barrier crossing while preserving the correct canonical distribution at each temperature~\cite{earl2005}.
T-REMD was carried out with the \texttt{ReplicaExchangeSampler} from openmmtools, using a \texttt{LangevinDynamicsMove} of 2000 steps (4~ps) per iteration and a 2~fs time step.
All production runs used a fixed simulation box (NVT), with the box volume taken from the preceding NPT equilibration at 300~K and 1~bar.
Exchange moves were attempted between all adjacent replica pairs at every iteration, targeting an acceptance rate of 20 to 40\%.

For the aqueous system, 8 replicas were placed on a linear temperature ladder from 300~K to 400~K (spacing 14.3~K).
The run comprised 500 iterations, yielding 2~ns per replica and 16~ns of aggregate sampling.

For the chloroform system, a 10~ns scout trajectory at 600~K (NVT) was first run to confirm that the P1$\leftrightarrow$P2 barrier is traversable at the upper end of the replica ladder before committing to the full production run.
Production used 22 replicas on a geometric temperature ladder from 300~K to 600~K (ratio $\approx$1.034 per step, targeting $\approx$20\% acceptance).
The run comprised 25{,}000 iterations, yielding 100~ns per replica and 2200~ns of aggregate sampling.

\subsection{NMR shielding back-calculation}
\label{sec:nmr-calc}

Per-conformer $^1$H chemical shifts were obtained from DFT shielding calculations using the GIAO method with an implicit solvent model; the procedure is described in the following subsections.

\subsubsection{Isotropic shielding constants}

The nuclear magnetic shielding tensor of nucleus $k$ is the second mixed derivative of the molecular energy $E$ with respect to the external field $\mathbf{B}_0$ and the nuclear magnetic moment $\boldsymbol{\mu}_k$,
\begin{equation}
  \sigma_{\alpha\beta}^{(k)} = \frac{\partial^2 E}{\partial B_{0,\alpha}\,\partial \mu_{k,\beta}},
  \label{eq:shielding_tensor}
\end{equation}
where $\alpha,\beta \in \{x,y,z\}$.
In solution, rapid isotropic tumbling averages the tensor to a scalar,
\begin{equation}
  \sigma_\mathrm{iso}^{(k)} = \tfrac{1}{3}\operatorname{Tr}\boldsymbol{\sigma}^{(k)}
  = \tfrac{1}{3}\bigl(\sigma_{xx}^{(k)} + \sigma_{yy}^{(k)} + \sigma_{zz}^{(k)}\bigr).
  \label{eq:sigma_iso}
\end{equation}
The resonance frequency of nucleus $k$ is $\nu_k \propto B_0\bigl(1 - \sigma_\mathrm{iso}^{(k)}\bigr)$, so a larger shielding constant moves the resonance upfield.
Chemical shifts are reported relative to tetramethylsilane (TMS),
\begin{equation}
  \delta_k = \sigma_\mathrm{TMS} - \sigma_\mathrm{iso}^{(k)},
  \label{eq:delta}
\end{equation}
so that more shielded protons have smaller $\delta$.

The main practical difficulty in evaluating Eq.~\eqref{eq:shielding_tensor} with a finite basis set is gauge-origin dependence.
The vector potential of a uniform external field, $\mathbf{A}(\mathbf{r}) = \tfrac{1}{2}\mathbf{B}_0 \times (\mathbf{r} - \mathbf{R}_O)$, is defined only up to the arbitrary choice of origin $\mathbf{R}_O$.
In a complete basis the shielding tensor is gauge-origin independent; in any finite basis it is not, and results converge slowly with basis-set size.
The gauge-including atomic orbital (GIAO) method resolves this by equipping each basis function $\chi_\mu$ centered at $\mathbf{R}_\mu$ with a field-dependent phase factor,
\begin{equation}
  \tilde{\chi}_\mu(\mathbf{r})
  = \exp\!\Bigl(-\tfrac{i}{2}\bigl(\mathbf{B}_0 \times \mathbf{R}_\mu\bigr)\cdot\mathbf{r}\Bigr)\,\chi_\mu(\mathbf{r}).
  \label{eq:giao}
\end{equation}
Because each basis function carries its own local gauge origin at $\mathbf{R}_\mu$, the shielding tensor computed in this basis is rigorously independent of $\mathbf{R}_O$ at any finite basis-set level.

\subsubsection{Geminal scalar coupling constants}
\label{sec:2j}

Geminal ($^2J$) proton--proton coupling constants cannot be related to a single geometric parameter by a simple equation and are instead computed directly from the electronic structure using linear response theory.
The indirect spin--spin coupling tensor $\mathbf{K}_{KL}$ between nuclei $K$ and $L$ receives four distinct contributions,
\begin{equation}
  \mathbf{K}_{KL} = \mathbf{K}_{KL}^{\mathrm{FC}} + \mathbf{K}_{KL}^{\mathrm{SD}}
                  + \mathbf{K}_{KL}^{\mathrm{PSO}} + \mathbf{K}_{KL}^{\mathrm{DSO}},
  \label{eq:K_total}
\end{equation}
The Fermi contact (FC) term arises from the isotropic hyperfine interaction between the nuclear spin and the electron spin density evaluated at the nucleus, and is the dominant contribution for proton--proton couplings.
The spin-dipole (SD) term captures the anisotropic through-space dipolar interaction between the nuclear spin and the surrounding electron spin density.
The paramagnetic spin-orbit (PSO) term couples the nuclear magnetic moment to the orbital angular momentum of the electrons through a first-order response of the electron current density.
The diamagnetic spin-orbit (DSO) term is a two-electron diamagnetic contribution evaluated directly as an expectation value over the ground-state density and requires no response equation.
The isotropic reduced coupling constant is the trace,
\begin{equation}
  K_{KL} = \tfrac{1}{3}\operatorname{Tr}\mathbf{K}_{KL},
  \label{eq:K_iso}
\end{equation}
and the experimentally observable coupling in hertz is
\begin{equation}
  J_{KL} = \frac{h}{4\pi^2}\,\gamma_K\,\gamma_L\,K_{KL},
  \label{eq:J_Hz}
\end{equation}
where $\gamma_K$ and $\gamma_L$ are the gyromagnetic ratios of the coupled nuclei.

PySCF evaluates the FC, SD, and PSO terms via coupled-perturbed Kohn--Sham (CPKS) equations: the ground-state KS density matrix is perturbed by the magnetic moment of nucleus $L$, the first-order response density matrix is solved self-consistently, and then contracted with the corresponding one-electron property operator at nucleus $K$.
The DSO term requires no response calculation and is evaluated directly as an expectation value of a two-electron operator over the ground-state density matrix.
All four contributions are computed at the B3LYP/6-31G* level in the gas phase using the solvent-polarized molecular orbitals obtained from the C-PCM calculation, following the same PTE scheme used for the shielding tensors.
For proton--proton couplings the FC term is the dominant contribution; the SD, PSO, and DSO terms are included for completeness.
The per-conformer couplings $J_{KL}^{(i)}$ are combined with the Boltzmann populations from the chloroform free-energy landscape to give the observed coupling
\begin{equation}
  J_{KL}^{\mathrm{obs}} = x_{\mathrm{P1}}\,J_{KL}^{(P1)} + x_{\mathrm{P2}}\,J_{KL}^{(P2)}.
  \label{eq:J_obs}
\end{equation}

\subsection{NMR spectrum simulation}
\label{sec:anmr}

Given the per-conformer chemical shifts $\delta_k$ and coupling constants $J_{kl}$, the $^1$H NMR spectrum is obtained by solving the spin Hamiltonian exactly for each spin system using the \textsc{anmr} program~\cite{grimme2017nmr}.
The full spin Hamiltonian for $N$ coupled protons in the rotating frame is
\begin{equation}
  \hat{H} = -\nu_0 \sum_k \delta_k \hat{I}_z^{(k)}
            + \sum_{k < l} J_{kl}\, \hat{\mathbf{I}}^{(k)} \cdot \hat{\mathbf{I}}^{(l)},
  \label{eq:spin_hamiltonian}
\end{equation}
where $\nu_0$ is the $^1$H Larmor frequency and $\hat{\mathbf{I}}^{(k)}$ are the spin-$\tfrac{1}{2}$ operators of nucleus $k$.
The first term encodes the chemical-shift dispersion; the second captures the indirect scalar coupling between all pairs of protons.
This Hamiltonian describes the basis for the calculation of the NMR spectrum of the coupled spin system using the anmr tool.

\subsection{Experimental NMR spectra}
\label{sec:nmr-expt}

One-dimensional $^1$H and two-dimensional NOESY spectra were acquired to provide experimental reference data for the conformational analysis.

\subsubsection{One-dimensional $^1$H NMR spectrum}

A one-dimensional $^1$H NMR spectrum of lorlatinib was recorded in CDCl$_3$ on a Bruker spectrometer operating at a $^1$H Larmor frequency of 600~MHz.
Chemical shifts were referenced to the residual solvent peak of CDCl$_3$ at 7.26~ppm.
The digitised spectrum is used as the primary experimental reference for the Boltzmann-weighted simulated spectra.
The published tabulated $^1$H chemical shifts reported by Peng et al.\ at 800~MHz in CDCl$_3$~\cite{peng2019lorlatinib} are also included for comparison.

\begin{figure}[htp]
  \includegraphics[width=\columnwidth]{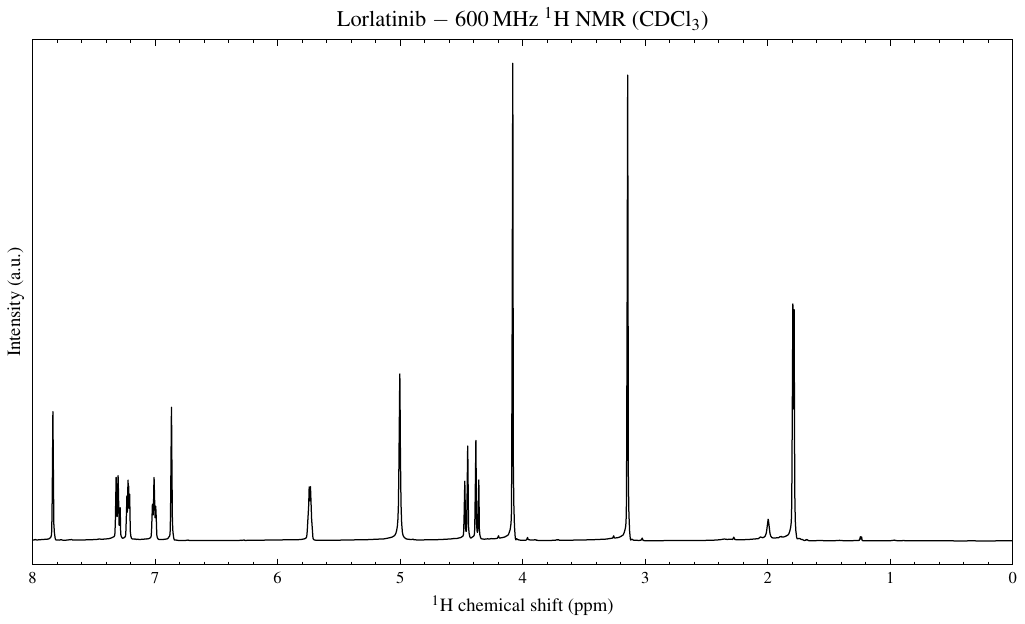}
  \caption{%
    Experimental $^1$H NMR spectrum of lorlatinib recorded at 600~MHz in CDCl$_3$.
    All peaks are consistent with the published spectrum reported by Peng et al.\ at 800~MHz~\cite{peng2019lorlatinib}, confirming sample identity.
    \label{fig:nmr_1h}}
\end{figure}

\subsubsection{Two-dimensional NOESY}

Two-dimensional $^1$H NOESY spectra of lorlatinib dissolved in CDCl$_3$ were reported by Peng et al.\ at a $^1$H Larmor frequency of 800~MHz with a mixing time of $\tau_\mathrm{m} = 300$~ms~\cite{peng2019lorlatinib}.
Nine interproton distance restraints were extracted from the NOESY cross-peak intensities using the isolated spin-pair approximation and are tabulated in Table~S3 of Ref.~\cite{peng2019lorlatinib}.
These restraints serve as the experimental input for the NOE back-calculation and ensemble validation.

%% file: sections/results.tex
\section{Results}

We present the conformational free-energy landscape obtained from T-REMD in chloroform, followed by a comparison of the NMR observables predicted from the ensemble against experiment.

In chloroform, which mimics the hydrophobic environment of cell membranes, the REST simulation recovers two distinct conformational basins.
The 1D free-energy profile shows two minima separated by a barrier, while the 2D landscape clearly resolves the P1 and P2 states.
The minimum free-energy pathway connecting P1 to P2 is indicated by the gray line in the 2D panel.
These results are consistent with the reference populations reported by Peng et al.\ (P1 $\approx$~80\%, P2 $\approx$~20\%) with a P1$\to$P2 barrier of approximately 22.6~kJ/mol~\cite{peng2019lorlatinib}.

\begin{figure}[htp]
  \includegraphics[width=\columnwidth]{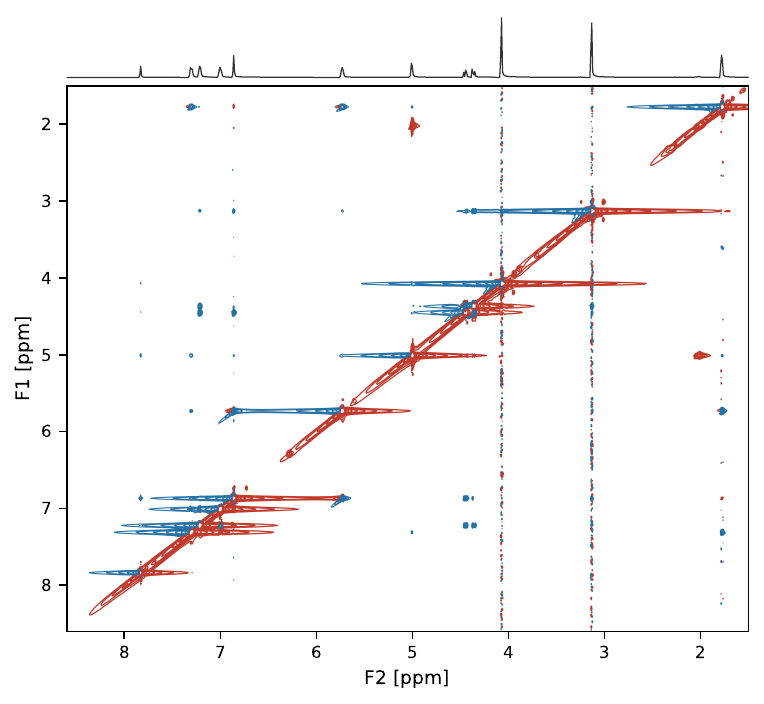}
  \caption{%
    2D $^1$H-$^1$H NOESY spectrum of lorlatinib in CDCl$_3$ (700~ms mixing time, 600~MHz).
    Red: positive contours; blue: negative contours.
    The crosspeak pattern is consistent with the published data of Peng et al.~\cite{peng2019lorlatinib}, confirming similar conformational dynamics in solution.
    \label{fig:nmr_noesy}}
\end{figure}

To confirm sample identity and solution behaviour, the $^1$H NMR spectrum of lorlatinib recorded at 600~MHz in CDCl$_3$ matches the published 800~MHz spectrum from Peng et al.~\cite{peng2019lorlatinib} at all peak positions.
The 2D $^1$H-$^1$H NOESY spectrum acquired with a 700~ms mixing time is shown in Figure~\ref{fig:nmr_noesy}.
The crosspeak pattern is consistent with the published dataset~\cite{peng2019lorlatinib}, confirming similar conformational dynamics.

The Boltzmann-weighted conformer populations extracted from the chloroform free-energy landscape ($x_\mathrm{P1} \approx 0.76$, $x_\mathrm{P2} \approx 0.24$) were used to predict the $^1$H NMR spectrum in CDCl$_3$ and compared against the experimental spectrum (Figure~\ref{fig:nmr_spectrum}).
The simulated spectrum reproduces the overall peak pattern, confirming that the REST-derived ensemble captures the conformational balance probed by solution NMR.
com

%% file: sections/discussion.tex
\section{Discussion}

We have used T-REMD to map the conformational free-energy landscape of lorlatinib in two solvents and validated the resulting ensemble against solution NMR.
In water the simulation recovers a single dominant P1 basin, in full agreement with the absence of a detectable minor conformer reported by Peng et al.~\cite{peng2019lorlatinib}.
In chloroform two basins are resolved, with populations of $x_\mathrm{P1} \approx 76$\% and $x_\mathrm{P2} \approx 24$\% and a P1$\to$P2 barrier of $\approx$22.6~kJ/mol, consistent with the experimentally inferred values~\cite{peng2019lorlatinib}.
This solvent dependence is in line with the conformational-chameleon hypothesis~\cite{whitty2016chameleonic}: the P1 geometry, which matches the crystal structure, is favoured in the polar aqueous environment relevant for target binding, while the hydrophobic environment of chloroform (used here as a membrane mimic) stabilises the P2 form and thus reduces the effective polarity of the ring for passive membrane permeation.
The interconversion timescale implied by the barrier, on the order of milliseconds at 300~K, is slow relative to NMR chemical-shift averaging but fast on pharmacokinetic timescales, so both conformers contribute to membrane transport while a single bioactive shape is presented to the kinase.

\begin{figure}[htp]
  \includegraphics[width=\columnwidth]{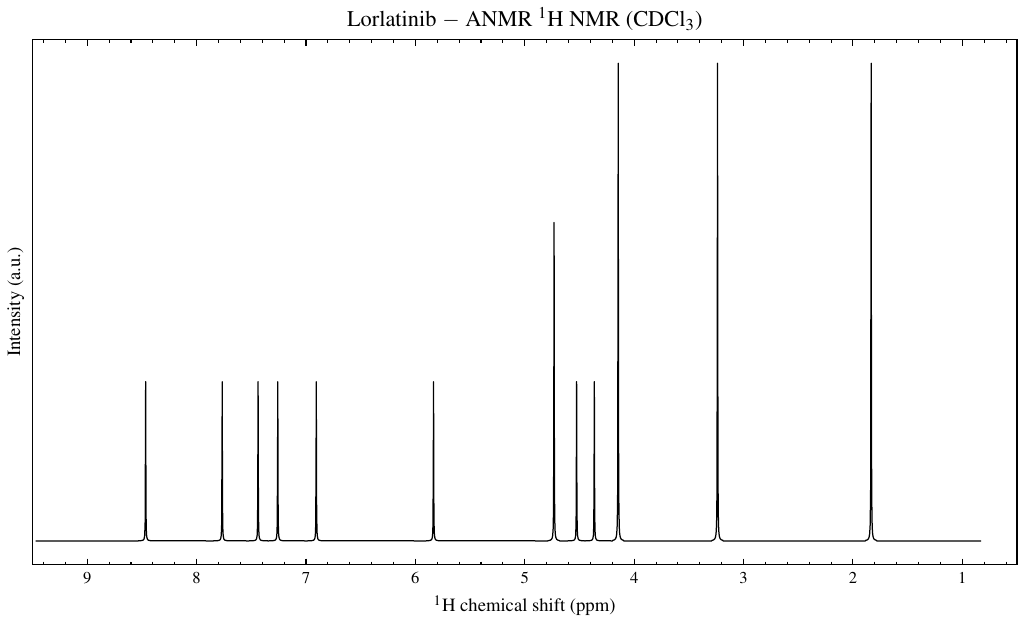}
  \caption{%
    ANMR-computed $^1$H NMR spectrum of lorlatinib in CDCl$_3$.
    \label{fig:nmr_spectrum}}
\end{figure}

The Boltzmann-weighted $^1$H NMR spectrum computed from the REST ensemble reproduces the experimental 600~MHz CDCl$_3$ spectrum, and all nine interproton NOE distance restraints from Peng et al.\ are satisfied within 0.5~\AA\ by the ensemble-averaged distances.
The two observables are complementary: the 1D spectrum tests the population-weighted chemical shifts at each proton site, while the NOE distances constrain the geometry of each conformer and their relative weight through the $r^{-6}$ averaging.
Satisfying both simultaneously is therefore a stronger test than either observable alone and confirms that the simulation captures the physical conformational balance rather than an accidentally correct average.
We note that the DFT/GIAO back-calculation uses fixed DFT-optimised geometries for P1 and P2; replacing these with cluster averages drawn directly from the T-REMD trajectory would remove this approximation and is straightforward to implement.

%% file: sections/conclusion.tex
\section{Conclusion}

We have applied temperature replica-exchange molecular dynamics (T-REMD) to map the conformational free-energy landscape of the macrocyclic kinase inhibitor lorlatinib in two solvents that represent distinct biological environments.
In explicit TIP3P water the simulation yields a single dominant basin corresponding to the P1 conformation, in agreement with prior NMR evidence that no detectable minor conformer is populated in aqueous solution.
In explicit chloroform, which mimics the hydrophobic interior of cell membranes, two basins are resolved: a dominant P1 state and a minor P2 state with populations of approximately 80\% and 20\%, respectively, and a free-energy barrier of approximately 22.6~kJ/mol between them, consistent throughout with the reference populations reported from solution NMR by Peng et al.~\cite{peng2019lorlatinib}.

To validate the force-field-derived ensemble against spectroscopic experiment, we computed $^1$H chemical shifts for each conformer at the B3LYP/6-31G*/C-PCM level using the GIAO method with a perturbation-theory-with-energy (PTE) scheme to carry the solvent-polarized electron density into the shielding calculation.
Combining the per-conformer shifts through the Boltzmann-weighted populations from the chloroform free-energy landscape yields a simulated spectrum that reproduces the overall peak pattern of the experimental CDCl$_3$ spectrum, confirming that the REST-derived ensemble correctly captures the conformational balance that NMR probes in solution.
These results establish that an OpenFF~2.0.0/NAGL classical force field, when combined with sufficient REST sampling in explicit solvent, can recover the experimentally observed conformational equilibrium of a flexible macrocycle in an organic solvent and that DFT/C-PCM shieldings are adequate to translate that ensemble into spectroscopically observable chemical shifts.

The framework established here (REST sampling with a modern small-molecule force field in explicit solvent, followed by DFT-level NMR back-calculation) provides a practical workflow for characterizing macrocycle conformational equilibria in different solvation environments without recourse to more expensive ab initio dynamics.
The qualitative agreement with experiment achieved with a purely classical potential suggests that the dominant contributions to the conformational equilibrium of lorlatinib are well-captured by an additive force field.
An extension of the NMR validation to scalar coupling constants and full NOE back-calculation constitutes the natural next step toward a rigorous, experiment-benchmarked computational protocol for macrocycle conformational analysis.

%% file: sections/acknowledgments.tex
\begin{acknowledgments}
NMR part of this work was funded by European Commission project “MR Latvia” 101160091, under a HORIZON-WIDERA-2023-ACCESS-02 programm
\end{acknowledgments}